\input harvmac
\def\np#1#2#3{Nucl. Phys. B {#1} (#2) #3}

\def\plb#1#2#3{Phys. Lett. B {#1} (#2) #3}

\def\physrev#1#2#3{Phys. Rev. D {#1} (#2) #3}

\def\G{{\cal G}}
\def\mH{{\cal M}_H}
%  draw box of size #1pt and line thickness #2pt
\def\drawbox#1#2{\hrule height#2pt 
        \hbox{\vrule width#2pt height#1pt \kern#1pt 
              \vrule width#2pt}
              \hrule height#2pt}
% Young tableaux

\def\Fund#1#2{\vcenter{\vbox{\drawbox{#1}{#2}}}}
\def\Asym#1#2{\vcenter{\vbox{\drawbox{#1}{#2}
              \kern-#2pt       % line up boxes
              \drawbox{#1}{#2}}}}
 
\def\fund{\Fund{6.5}{0.4}}
\def\asym{\Asym{6.5}{0.4}}
\batchmode
  \font\bbbfont=msbm10
\errorstopmode
\newif\ifamsf\amsftrue
\ifx\bbbfont\nullfont
  \amsffalse
\fi
\ifamsf
\def\IR{\hbox{\bbbfont R}} 
\def\IC{\hbox{\bbbfont C}}

\def\IZ{\hbox{\bbbfont Z}}
\def\IF{\hbox{\bbbfont F}}
\def\IP{\hbox{\bbbfont P}}
\else
\def\IR{\relax{\rm I\kern-.18em R}}
\def\IC{\relax{\rm {\bf C}}}
\def\IZ{\relax\ifmmode\hbox{Z\kern-.4em Z}\else{Z\kern-.4em Z}\fi}
\def\IF{\relax{\rm I\kern-.18em F}}
\def\IP{\relax{\rm I\kern-.18em P}}
\fi
\def\N#1#2{${\cal N}=(#1, #2)$}
\def\W{\cal W}
\def\R{{\cal R}}
\def\P{{\cal P}}
\def\C{{\cal C}}
\def\FI{Fayet-Iliopoulos}
\def\lfm#1{\medskip\noindent\item{#1}}
\lref\Betal{M. Berkooz, R.G. Leigh, J. Polchinski, J. Schwarz,
N. Seiberg, and E. Witten,  hep-th/9605184, \np{475}{1996}{115}.}
\lref\obrane{K. Intriligator,  hep-th/9702038, \np{496}{1997}{177}.}
\lref\obraneii{J.D. Blum and K. Intriligator, hep-th/9705044.}
\lref\snewt{N. Seiberg, hep-th/9705221.}
\lref\IMS{K. Intriligator, D.R. Morrison, and N. Seiberg,
hep-th/9702198, Nucl. Phys. B to appear.}
\lref\mrdgs{M.R. Douglas, hep-th/9612126.}
\lref\berkdou{M. Berkooz and M. R. Douglas, hep-th/9610236.}
\lref\MRDrev{M.R. Douglas, hep-th/9707228.}
\lref\aspgro{P. S. Aspinwall and M. Gross, hep-th/9605131, 
\plb{387}{1996}{735}.}
\lref\aspmor{P.S. Aspinwall and D.R. Morrison, hep-th/9705104.}
\lref\bershetal{M. Bershadsky,  K. Intriligator, S. Kachru, 
D.R. Morrison, V. Sadov, and C. Vafa, hep-th/9605200,
\np{481}{1996}{215}.}
\lref\naka{Nakajima, J. Diff. Geom. 40 (1990) 105; Duke Math 76 (1994)
365.}
\lref\ismir{K. Intriligator and N. Seiberg, hep-th/9607207,
\plb{387}{1996}{513}.}
\lref\perraj{E. Perevalov and G. Rajesh, hep-th/9706005.}
\lref\hanwit{A. Hanany and E. Witten, hep-th/9611230, 
\np{492}{1997}{152}.}
\lref\harmoo{J. Harvey and G. Moore, hep-th/9609017.}
\lref\natisav{N. Seiberg and S. Sethi, hep-th/9708085.}
\lref\asp{P.S. Aspinwall, hep-th/9612108, \np{496}{1997}{149}.}
\lref\dm{M.R. Douglas and G. Moore, /9603167.}
\lref\tadpole{J.D. Blum and K. Intriligator, hep-th/9705030.}
\lref\bfss{T. Banks, W. Fischler, S.H. Shenker, L. Susskind,
hep-th/9610043, \physrev{55}{1997}{5112}.}
\lref\brs{M. Berkooz, M. Rozali, N. Seiberg, hep-th/9704089.}
\lref\bervaf{M. Bershadsky and C. Vafa, hep-th/9703167.}
\lref\wsem{E. Witten, seminar at Aspen Center for Physics, Aug. '97.}
\Title{hep-th/9708117, IASSNS-HEP-97/95, UCSD/PTH-22}
{\vbox{\centerline{New String Theories in Six Dimensions via}
\centerline{Branes at Orbifold Singularities}}}
\medskip
\centerline{Kenneth Intriligator}
\vglue .5cm
\centerline{School of Natural Sciences}
\centerline{Institute for Advanced Study}
\centerline{Princeton, NJ 08540, USA}
\vglue .25cm
\centerline{and}
\vglue .25cm
\centerline{UCSD Physics Department\footnote{${}^*$}{new address}}
\centerline{9500 Gilman Drive}
\centerline{La Jolla, CA 92093}

\bigskip
\noindent

We present several classes of new 6d string theories which arise via
branes at orbifold singularities.  They have compact moduli spaces,
associated with tensor multiplets, given by Weyl alcoves of
non-Abelian groups.  We discuss T-duality and Matrix model
applications upon compactification.

\Date{8/97}                                   
%\draftmode        

\newsec{Introduction}

It was recently pointed out in \snewt\ that new 6d theories, which
include stringy excitations but without gravity, can be obtained in
the world-volume of five-branes by taking $g_s\rightarrow 0$ with
$M_s$ held fixed.  Four different classes were obtained in \snewt:

\lfm{(iia)} Theories with \N{1}{1}\ supersymmetry, 
which are obtained in type IIB five-branes or, alternatively \wsem,
via type IIA with a $\IC ^2/\Gamma _G$ ALE singularity
\foot{We label $\Gamma _G\subset
SU(2)$ using the well-known correspondence with the simply-laced
groups $G=A_r,\, D_r,\, E_{6,7,8}$.}.

\lfm{(iib)} Theories with
\N{2}{0}\ supersymmetry, which are obtained in the world-volume of
type IIA (or M-theory) five branes or, alternatively \wsem, via type
IIB with a $\IC ^2/\Gamma _G$ singularity.

\lfm{(o)} Theories with \N{1}{0}
supersymmetry in the world-volume of $SO(32)$ heterotic
small-instantons or type I five-branes.

\lfm{(e)} Theories with \N{1}{0}
supersymmetry in the world-volume of $E_8$ small instantons.

\noindent 
The $(o)$ theory has a global $SO(32)$ symmetry and the $(e)$ theory
has a global $E_8\times E_8$ symmetry.

The infrared limit of these theories, with energies small compared to
$M_s$, appear to be local quantum field theories.  In the $(iib)$ and
$(e)$ cases these are non-trivial, interacting, RG fixed points, while
the $(iia)$ and $(o)$ cases are IR free.  Despite their different IR
behavior, upon compactification to five-dimensions on a circle, $T$
duality exchanges the $(iia) \leftrightarrow (iib)$ and
$(o)\leftrightarrow (e)$ theories.  Thus the full theories are {\it
not} local quantum field theories \snewt.

In this paper, we discuss new 6d \N{1}{0}\ theories associated
with type II or heterotic five-branes at orbifold singularities
in the orthogonal four dimensions.  As in \snewt, we take $g_s
\rightarrow 0$ with $M_s$ fixed.  The fact that new theories could
be thus obtained was also mentioned during the course of this work in
a footnote in \MRDrev.  It was there pointed out that one could have a
general, Ricci-flat, non-compact manifold ${\cal M}_4$ in the remaining
four directions, giving theories which, in principle, could depend on
the uncountably infinite parameters needed to specify ${\cal M}_4$.
However, as in \natisav, we expect most of these parameters are
irrelevant in the $g_s\rightarrow 0$ and that only the $\IC ^2/\Gamma
_G$ singularity type matters.  Clearly the singularity itself can not
be ignored; indeed, it breaks the supersymmetry of the $(iia)$ or
$(iib)$ theories to \N{1}{0} supersymmetry.

The 6d string theories which we present have compact ``Coulomb
branches,'' associated with expectation values the scalar components
of 6d \N{1}{0}\ tensor multiplets, which are the ``Coxeter boxes''
(also referred to as the ``Weyl alcove'') of non-Abelian groups.
For any group $\G$ of rank $r$, the Coxeter box is a compact
subspace of $\IR ^r$ given by all $\vec \Phi \in \IR ^r$ which satisfy
\eqn\cbox{\vec \alpha _\mu \cdot \vec \Phi +
M_s^2 \delta _{\mu 0}\geq 0, \qquad \mu =0\dots r,} where $\vec
\alpha _\mu$ are the simple roots, including the extended root $\mu
=0$, with $\sum _{\mu =0}^rn_\mu \vec \alpha _\mu =0$ ($n_\mu$ are the
Dynkin indices).  The $\mu \neq 0$ conditions in \cbox\ give the
non-compact Weyl chamber $\IR ^r/\W _G$, where $\W _G$ is the
Weyl-group.  Including the $\mu =0$ condition gives the Coxeter box
$\IR ^r/{\cal C} _\G\cong (S^1)^r/\W _\G$, where the Coxeter group
${\cal C}_\G$ includes translations in the root lattice of $\G$.
Compact Coxeter box moduli spaces, of size $R^{-1}$, also arise via
Wilson loops upon reducing a $\G$ gauge theory on a circle of radius
$R$.  We have written the size of the Coxeter box
\cbox\ as $M_s^2$ because here it will be.

Coxeter boxes already appear in the theories $(iib)$ and $(e)$
mentioned above.  Part of the moduli space of the $(iib)$ theory
obtained from $K$ parallel five-branes is the $U(K)$ Coxeter box of
size $M_s^2$.  The $(iib)$ theory obtained from type IIB string theory
on a $\IC ^2/\Gamma _G$ ALE singularity has, as part of its moduli
space, the Coxeter box of size $M_s^2$ of the corresponding $ADE$
group $G$.  The $(e)$ theory obtained from $K$ small $E_8\times
E_8$ instanton five-branes has the Coxeter box, again of size $M_s^2$,
for $Sp(K)$ as its Coulomb branch.

We will simply note some basic features of the new 6d string theories,
saving a more detailed analysis for further study.  In the next
section, we discuss theories associated with type IIB NS five-branes
at orbifold singularities.  The tensor multiplet moduli live on the
Coxeter box of the simply laced group $G$ associated with the
singularity.  In sect. 3 we discuss theories associated with $SO(32)$
heterotic or type I branes at orbifold singularities.  In these
examples, the tensor multiplet moduli can live in the Coxeter box of a
non-simply-laced subgroup of $G$.  In sect. 4, we discuss theories
associated with $E_8\times E_8$ branes at orbifold singularities.  In
sect. 5, we discuss $T$ duality upon compactification.  Finally, in
sect. 6, we discuss applications of the theories to providing a
definition of $M$ theory on $(ALE)\times T^5\times
\IR ^{1,1}$ and $M$ theory on $(ALE)\times (T^5/\IZ _2)\times \IR ^{1,1}$.

\newsec{Type IIB branes at a $\IC ^2/\Gamma _G$ orbifold singularity}

For our first class of examples, consider $K$ parallel type IIB NS
five-branes at a $\IC ^2/\Gamma _G$ orbifold singularity in the
transverse directions.  Having five-branes but no ALE singularity
would lead to a $(iia)$ theory of \snewt.  Having the ALE singularity
but no five-branes would lead to a $(iib)$ theory of \snewt.  Putting
the two situations together leads to new
\N{1}{0}\ string theories, whose field theory infra-red limit 
was discussed in \obraneii.

As discussed in \obraneii, the  
\N{1}{0} theory has gauge group 
\eqn\iibggi{\prod _{\mu =0}^r  U
(Kn_\mu ),} 
with matter multiplets in the representations $\half
\oplus _{\mu \nu} a_{\mu \nu}(\fund _\mu ,\overline{\fund} _\nu)$.  
In addition, there are $r\equiv\hbox{rank}\, G$ hyper-multiplets and
tensor multiplets (which would give $r$ \N{2}{0}\ matter multiplets
for the theory with no five-branes).  $r$ of the $U(1)$ factors in
\iibggi\ have charged matter and are thus anomalous in 6d.  As in 
\refs{\dm, \Betal}, this means that these $U(1)$ factors are spontaneously
broken; they pair with the $r$ hyper-multiplets mentioned above to get
a mass.  The massless, unbroken gauge group is thus
\eqn\iibgg{U(1)\times \prod _{\mu =0}^r SU(Kn_\mu),}
with the $U(1)$ factor decoupled, with no charged matter.  
Although the $U(1)$ factors in \iibggi\ are massive, their $D$ term
equations still constrain the moduli space. Supersymmetry implies that 
the expectation values of the $r$ hyper-multiplets involved in the
$U(1)$ anomaly cancelation appear as \FI\ terms in these constraints
\dm; these are the ALE blowing-up modes, which enter as background
parameters in the 6d theory.

Taking $g_s\rightarrow 0$ with $M_s$ fixed, the tensor multiplet
moduli space is the Coxeter box \cbox\ of the corresponding $ADE$
group $G$.  This can be seen starting from the $(iib)$ theory
associated with the ALE space and no branes.

Using results found in
\refs{\obrane,
\obraneii} via anomalies, the effective gauge coupling of the $r+1$
gauge groups on the Coulomb branch can be written as
\eqn\geffc{g_\mu ^{-2}(\vec \Phi)=\vec \alpha
_\mu \cdot \vec \Phi +M_s^2\delta _{\mu 0}} where, as in \cbox, the
$\vec \alpha _\mu$ are the simple and extended roots of the ADE group
$G$ associated with the singularity.  Using $\vec \alpha _\mu \cdot
\vec \alpha _\mu=\widetilde C_{\mu \nu}$, 
the extended Cartan matrix of $G$, the couplings in \geffc\ cancel the
reducible $\widetilde C_{\mu \nu}\tr F_\mu ^2 \tr F_\nu ^2$ anomaly
terms found in \refs{\obrane, \obraneii}.  We see that, as required,
all $g_\mu ^{-2}\geq 0$ over the entire Coulomb box \cbox, with the
various $g_\mu ^{-2}=0$ along the boundaries of the Coulomb box.  The
``Landau pole'' mentioned in \refs{\obrane, \obraneii} has been
eliminated by the compactness of the Coulomb branch for finite $M_s$.

There is a Higgs mode of the theory corresponding to moving the $K$
five-branes away from the $X_G\cong  \IC ^2/\Gamma _G$ ALE space.
This Higgs branch moduli space is $\mH\cong (X_G)^K/S_K$, as expected,
with \iibgg\ broken to the diagonal $U(K)_D$ away from the origin (or
with non-zero \FI\ parameters).  This $U(K)_D$ theory is the $(iia)$
theory of the branes away from the singularity, with gauge coupling
$g_D^{-2}= \sum _{\mu =0}^r n_\mu g_\mu ^{-2}=M_s^2$ as expected.
The low energy theory has an enhanced, accidental
\N{1}{1}\ supersymmetry which is not respected by the massive field theory
and stringy modes.

There are also interesting new 6d theories associated with type $IIA$
NS 5-branes at orbifold singularities, which require further
understanding.  For the case of $K$ branes at a $\IC ^2/\IZ _M$
singularity, the 6d theory could be the same theory as that of $M$
type $IIB$ branes at a $\IC ^2/\IZ _K$ singularity (up to a decoupled
tensor multiplet in the former and vector multiplet in the latter).

\newsec{New theories from $SO(32)$ branes at ALE singularities}

Our next class of new 6d string theories with \N{1}{0}\ supersymmetry
arise from $SO(32)$ heterotic or type I 5-branes at $\IC ^2/\Gamma _G$
orbifold singularities.  The low energy limit of these theories was
discussed in \refs{\obrane, \tadpole, \obraneii} and also, via
F-theory, in \refs{\asp, \aspmor}.  The gauge group is  
\eqn\sigg{\prod _{\mu \in \R}Sp(v_\mu )\times \prod _{\mu \in
\P}SO(v_\mu )\times \prod _{\mu \in \C}U(v_\mu ),}
where the nodes of the extended $G$ Dynkin diagram have been grouped
into the sets $\R ,\ \P ,\ \C ,\ \overline \C$ discussed in detail in
\obraneii.  As in the discussion 
following \iibggi, the overall $U(1)$ factor in each $U(v_\mu)$ is
anomalous and thus pairs with a hyper-multiplet to get a mass.

The tensor multiplet structure is related to the Coxeter box of the
corresponding simply-laced group $G$, but modded out by a $\IZ _2$
action $*$ which takes $\C
\leftrightarrow \overline \C$.  From the analysis in \refs{\tadpole,
\obraneii}, the result is that the tensor
multiplets for a $\IC ^2/\Gamma _G$ singularity live in the Coxeter
box of $H\subset G$ with $G\rightarrow H$ as
\eqn\GGR{\matrix{&SU(2P)&\rightarrow &Sp(P)\cr
&SO(4P+2)&\rightarrow &SO(4P+1)\cr &SO(4P)&\rightarrow &SO(4P)\cr &E_6
&\rightarrow &F_4 \cr &E_7&\rightarrow & E_7\cr &E_8&\rightarrow
&E_8.}}

The operation in \GGR\ is the same modding out which appeared in the
description of \refs{\aspgro, \bershetal} for obtaining composite
gauge invariance with non-simply-laced gauge groups.  Although it is
outside of the focus of this work, we note that the hyper-Kahler
quotient construction of \refs{\obrane, \obraneii} for the moduli
space of $SO(N)$ instantons on ALE spaces suggests an interesting
analog of the results of Nakajima.  Briefly put, Nakajima \naka\
showed that $\widehat G _N$ affine Lie algebras arise in analyzing the
moduli space of $U(N)$ instantons on $\IC ^2/\Gamma _G$.  Similarly,
we expect $\widehat H _N$ affine Lie algebras to arise in analyzing
the moduli space of $SO(N)$ instantons on $\IC ^2/\Gamma _G$, with
$G\rightarrow H$ as in \GGR.  The results of \naka\ find physical
application, for example in \harmoo, in showing that simply-laced
composite gauge invariance is properly represented on massive modes.
The conjectured appearance of $\widehat H_N$ affine Lie algebras could
find similar application in compactifications with non-simply-laced
composite gauge invariance.

Other 6d theories can be obtained by making use of the fact, as in
\Betal, that the gauge group of the heterotic or type I theory 
is actually $Spin(32)/\IZ _2$.  The low-energy limit of these string
theories in the case of $\IC ^2/\IZ _{2P}$ singularities was discussed
in \obrane, where it was (sloppily) referred to as the case without
vector structure.  The result is a theory based on the ``type I5
quiver diagrams'' of \dm, with gauge group 
\eqn\novgg{\prod _{i=1}^P SU(v_\mu)}
and tensor multiplets which live in the Coxeter box, of size $M_s^2$,
of $Sp(P-1)$.  For the simplest example, $\IC ^2/\IZ _2$, the low
energy theory is $SU(2K)$ with two matter fields in the $\asym$ and
sixteen in the $\fund$ and no tensor multiplet.

\newsec{New theories from $E_8\times E_8$ branes at orbifold singularities}

Our next class of new 6d string theories with \N{1}{0}\ supersymmetry
arise via $E_8\times E_8$ 5-branes at orbifold singularities in the
$g_s\rightarrow 0$ with $M_s$ fixed limit.  The gauge group and number
of tensor multiplets associated with point-like $E_8$ instantons at
ADE orbifold singularities was obtained via F-theory in \aspmor.  We
take this opportunity to briefly spell out the massless matter content
of these theories, which we determine from the results of \aspmor\
combined with anomaly considerations, as it was not presented in
\aspmor.  First, the
irreducible $\tr F^4$ gauge anomalies must vanish; remaining reducible
anomalies must then be canceled by coupling to the tensor multiplets.
In addition, as discussed in \bervaf, a $\pi _6$ anomaly restricts
$SU(2)$ to have $n_2=4$ mod 6, $SU(3)$ to have $n_3=0$ mod 6, and
$G_2$ to have $n_7=1$ mod 3.  A further general condition is
\eqn\anomatch{n_H-n_V+29n_T=30K+r,}
where $n_H$ is the total number of hyper-multiplets, $n_V$ is the
total number of vector multiplets, $n_T$ is the number of tensor
multiplets, $K$ is the number of small instantons or five-branes, and
$r\equiv$rank$G$ is the number of ALE blowing-up modes.  The condition
\anomatch\ is a 6d analog of a 't Hooft anomaly matching condition for
the gravitational anomaly.

The theory $(e)$ for $K$ $E_8\times E_8$ five-branes and no singularity has
a Coulomb branch with $n_T=K$ tensor multiplets and no vector-multiplet
gauge group.  Putting the $K$ 5-branes at a $\IC ^2/\IZ _M$ singularity, 
with $K\geq 2M$, the result of \aspmor\ is that there is a Coulomb branch,
again with $n_T=K$ tensor multiplets, but with new gauge fields, with gauge
group
\eqn\agg{SU(2)\otimes SU(3)\otimes \cdots \otimes SU(M-1)\otimes
SU(M)^{\otimes (K-2M+1)}\otimes SU(M-1)\otimes \cdots \otimes SU(2).}
The massless matter content consists of bi-fundamentals charged under each
neighboring pair of gauge groups in
\agg\ as well as an extra fundamental flavor for each of the two $SU(2)$s
at the ends and for each of the two $SU(M)$s at the end of the string
of $SU(M)s$.  As remarked in \aspmor, the gauge group in \agg\ agrees
(up to replacing the $SU(n)$ with $U(n)$) with that of \refs{\ismir,
\hanwit} which is mirror dual in three dimensions to $U(M)$ gauge
theory with $K$ flavors; the above hyper-multiplet content also agrees
with that of \refs{\ismir, \hanwit}.  The theory with this gauge group
and matter content is properly free of gauge anomalies (making use of
couplings to $K-3$ of the tensor multiplets to cancel the reducible
gauge anomalies).  

The theory with the above gauge group and matter content properly has
a $K+M-1$ dimensional Higgs branch, with the gauge group generically
completely broken. $M-1$ of the Higgs-branch moduli correspond to the
blowing-up modes of the $\IC ^2/\IZ _M$ orbifold.  The remaining $K$
dimensions is the $K$-fold symmetric product of the ALE space with
those $M-1$ moduli, corresponding to the locations of the $K$
identical, point-like instantons on the ALE space.  For generic values
of these moduli, the 5-branes are away from any singularity and there
are no vector-multiplets; the gauge symmetry \agg\ is unHiggsed when the
moduli are tuned, corresponding to putting the 5-branes on the singularity.

For $K=6$ five-branes at a $G=D_4$ singularity, the result of \aspmor\
is that the gauge group is $SU(2)\times G_2\times SU(2)$ with $n_T=6$
tensor multiplets.  The matter content is determined by anomaly
considerations to be $\half {\bf (2,1,1)}\oplus \half {\bf (2,7,1)}\oplus
\half {\bf (1,7,2)}\oplus \half {\bf (1,1,2)}\oplus 2{\bf (1,7,1)}$.  
This theory has a 10 dimensional Higgs branch, with the gauge group
generically completely broken, corresponding to the location of the
six point-like instantons on the ALE space and its four blowing-up
modes.  Giving an expectation value to a matter fields in the ${\bf
(1,7,1)}$ corresponds to smoothing the $D_4$ singularity to an $A_2$
singularity.

For $K\geq 7$ five-branes at a $G=D_4$ singularity, the result of
\aspmor\ is gauge group $SU(2)\times G_2\times SO(8)^{K-7}\otimes
G_2\otimes SU(2)$ with $n_T=2K-7$.  The matter content is determined
by anomaly considerations to be $\half {\bf (2,1)}\oplus \half {\bf
(2,7)}$ for each $SU(2)\times G_2$ pair and no other matter fields.

For $E_6$, the result of \aspmor\ is $n_T=4K-22$, with gauge group
$SU(2)\times G_2\times F_4\times G_2\times SU(2)$ for $K=8$ and gauge
group $SU(2)\times G_2\times F_4\times SU(3)\times (E_6\times
SU(3))^{K-9}\times F_4\times G_2\times SU(2)$ for $K>8$.  The matter
content is determined by anomaly considerations to consist, as above,
of the minimal $SU(2)\times G_2$ matter $\half {\bf (2,1)}\oplus \half
{\bf (2,7)}$ in each pair of $SU(2)\times G_2$.  For $K=8$ the $F_4$
has a single matter field in the ${\bf 26}$ (giving it an expectation
value breaks $F_4\rightarrow SO(9)\rightarrow SO(8)$, corresponding to
smoothing the singularity from $E_6\rightarrow D_5\rightarrow D_4$).
For $K>8$ each $SU(2)\times G_2$ pair has the same minimal matter
content as above, and there is no other matter.

For $K\geq 10$ five-branes at a $E_7$ singularity, the result of
\aspmor\ is $(SU(2)\times G_2)^4\times F_4^2\times E_7\times
(SU(2)\times SO(7)\times SU(2)\times E_7)^{K-10}$ with $n_T=6K-40$.
Each $SU(2)\times G_2$ factor has the minimal matter appearing above.
Each $SU(2)\times SO(7)\times SU(2)\times E_7$ factor has matter $\half {\bf
(2,8,1,1)}\oplus \half {\bf (1,8,2,1)}$.  There is no other matter.

For $K\geq 10$ five-branes at a $E_8$ singularity, the result of
\aspmor\ is gauge group $E_8^{(K-9)}\times F_4^{(K-8)}\times
(SU(2)\times G_2)^{2K-16}$ with $n_T=12K-96$.  Each $SU(2)\times G_2$
factor has the minimal matter content appearing above and there is no
other matter.

The result of \aspmor\ for $K=2m+6$ five-branes at a $D_{m+4}$
singularity is $n_T=2K-6$ and gauge group $SU(2)\times G_2\times
SO(9)\times SO(3)\times SO(11)\times SO(5)\times \cdots \times
SO(2m+5)\times SO(2m-1)\times SO(2m+7)\times SO(2m-1)\times \cdots
\times SO(9)\times G_2\times SU(2)$.  For $K>2m+8$ five-branes at a
$D_{m+1}$ singularity, \aspmor\ again find $n_T=2K-6$ and, in addition
to the gauge group factors for $K=2m+6$, $Sp(m)\times (SO(2m+8)\times
Sp(m))^{(K-2m-8)}\times SO(2m+7)$.  For $m>1$, we were not able to
find a solution for matter content which is compatible with anomaly
considerations and these gauge groups, though perhaps one does
exist\foot{ {\bf Note added} (in revised version, 9/3/97): There is a
slight modification of the above gauge groups for which there is a
matter content which {\it is} nicely compatible with all of the
anomaly considerations.  For $K=2m+6$ five-branes at a $D_{m+4}$
singularity, with $n_T=2K-6$ as in \aspmor, the modified gauge group
is $SU(2)\times G_2\times
SO(9)\times Sp(1)\times SO(11)\times Sp(2)\times \cdots \times
SO(2m+5)\times Sp(m-1)\times SO(2m+7)\times Sp(m-1)\times \cdots
\times SO(9)\times G_2\times SU(2)$.  The matter 
content which satisfies all of the anomaly equations is given by the
minimal $\half ({\bf (2,1)\oplus (2,7)})$ in each $SU(2)\times G_2$
factor and a half-hypermultiplet bi-fundamental charged under each
neighboring $SO$ and $Sp$, i.e. a $\half {\bf (2k+7,2k)}$ under each
neighboring $SO(2k+7)\times Sp(k)$ and a $\half {\bf (2k,2k+9)}$ under
each neighboring $Sp(k)\times SO(2k+9)$.  In addition, the middle
$SO(2m+7)$ gauge group has a hypermultiplet in the ${\bf 2m+7}$ which
is uncharged under the other gauge groups.  For the cases $m=2,3$,
where the gauge group agrees with that of \aspmor\ (as $Sp(1)\cong
SO(3)$ and $Sp(2)\cong SO(5)$), this matter content was first worked
out by G. Rajesh.  I am very grateful for his correspondence on the
$m=2,3$ cases, which helped to inspire the above modified gauge groups
and matter content for $m>3$.  I also thank P.S. Aspinwall and
D.R. Morrison for helpful correspondence on these issues.  A similar
modification of the gauge group and matter content applies for
$K>2m+8$.}.

\newsec{Compactification and $T$ duality}

It is natural to expect that, upon compactification on a circle, the
new theories associated with five-branes at singularities are related
by $T$ duality, generalizing that of \snewt\ between
$(iia)\leftrightarrow (iib)$ and $(o)\leftrightarrow (e)$.  As in
\snewt, this can be put to a simple test.

Upon compactifying on a circle, both the Cartan of the 6d gauge group
and the 6d tensor multiplets lead to 5d $U(1)$ gauge fields with
scalar moduli. The number of 5d scalar moduli is thus $r_V+n_T$, where
$r_V$ is the rank of the 6d vector multiplet gauge group and $n_T$ is
the number of 6d tensor multiplets.  Two 6d theories related by $T$
duality must thus have $r_V+n_T=\widetilde r_V+\widetilde n_T$.  More
precisely, tensor multiplets in 6d have a compact ``Coulomb branch,''
with the scalar moduli living on a box of size $M_s^2$.  Upon reducing
to 5d and rescaling the modulus to have dimension one, it lives on a
box of size $M_s^2R$.  On the other hand, reducing a 6d vector
multiplet to 5d leads to a scalar modulus which lives on a box of size
$R^{-1}$.  Because $T$ duality relates a theory compactified on a
circle of radius $R$ to another theory compactified on a circle of
radius $\widetilde R\equiv (M_s^2R)^{-1}$, it exchanges 5d moduli
associated with 6d tensor multiplets with those associated with 6d
vector multiplets.  Thus $T$ dual theories must satisfy the stronger
conditions $\widetilde r_V=n_T$ and $\widetilde n_T=r_V$.

This can be thought of as a reason why, as we have seen, the Coulomb
branch of 6d tensor multiplets is the Coxeter box of a non-Abelian
group.  Compactifying on a circle, there should be a $T$ dual
theory where these moduli {\it do} arise from a gauge theory with that
gauge group.

For example, the vector multiplets of the $(iia)$ theory compactified
on a circle of radius $R$ and the tensor multiplets of the $(iib)$
theory compactified on a circle of radius $\widetilde R\equiv
(M_s^2R)^{-1}$ both lead to moduli living on a Coxeter box of size
$R^{-1}$, compatible with their equivalence \snewt.  Similarly, both
the $SO(32)$ theory $(o)$, compactified on a circle of radius $R$,
with Wilson lines which break it to $SO(16)\times SO(16)$, and the
theory $(e)$ on a circle of radius $\widetilde R\equiv (M_s^2R)^{-1}$
lead to a 5d moduli space which is the Coxeter box of $Sp(K)$, of size
$R^{-1}$.

The theories associated with $SO(32)$ and $E_8\times E_8$ branes at
$\IC ^2/\Gamma _G$ singularities do satisfy the condition
$r_V+n_T=\widetilde r_V+\widetilde n_T$.  Indeed, as also noted in
\perraj, in both cases, $r_V+n_T=C_2(G)K-|G|$, where $C_2(G)$ is the
dual Coxeter number of the $ADE$ group $G$ and $|G|$ is its
dimension\foot{As also noted in \perraj, this agrees with the
dimension (in hyper-multiplet units) of the moduli space of $K$ $G$
instantons on $K3$.  Duality between the heterotic theory on $T^3$ and
$M$ theory on $K3$ suggests that the quantum-corrected Coulomb branch
for the theory compactified to 3d on a $T^3$ {\it actually is} the
moduli space of $K$ $G$ instantons on $K3$.  Similarly, compactifying
the theory of sect. 2 associated with type II branes at orbifold
singularities, the dimension of the Coulomb branch is $C_2(G)K$.
Duality between type II on a $T^3$ and $M$ theory on $T^4$ suggests
that the quantum-corrected Coulomb branch for the theory compactified
to 3d on a $T^3$ is the moduli space of $K$ $G$ instantons on $T^4$.}.
On the other hand, the two theories do not satisfy the stronger
conditions $\widetilde r_V=n_T$ and $\widetilde n_T=r_V$.  It is not
presently known how this failure should be interpreted or resolved.

\newsec{Matrix Model Applications of the Theories.}

Following \bfss, it was suggested in \mrdgs\ that a M(atrix)
description of $M$ theory on $X_G\times \IR ^{6,1}$, where $X_G$ is an
ALE space asymptotic to $\IC ^2/\Gamma _G$, is given by quantum
mechanics with $8$ supersymmetries and gauge group $\prod _{\mu =0}^r
U(v_\mu)$ with matter $\half \oplus _{\mu \nu =0}^r a_{\mu \nu}(\fund
_\mu , \overline{\fund} _{\nu})$.  The (classical) moduli space of
vacua of this theory for $v_\mu =Kn_\mu$ is \foot{This is the moduli
space for generic Higgs expectation values.  There is a larger Coulomb
branch, of dimension $5KC_2(G)$, at the origin.} $(X_G\times \IR
^5)^K/S_K$, corresponding to the location of $K$ identical zero branes
in the light-cone $X_G\times \IR ^5$.  We propose a slight variant of
this conjecture.

Now consider $M$ theory on $X_G\times T^5\times R^{1,1}$.  Following
\snewt, it is expected\foot{I thank N. Seiberg for suggesting this.} 
that a definition of this theory is given by compactifying the new 6d
theory of sect. 2 on a $\widehat T^5$.  As in \snewt, there are 25
compactification parameters living in $SO(5,5,\IZ )\backslash SO(5,5)/
(SO(5)\times SO(5))$.  Taking a rectangular torus with no $B$ field,
$\widehat T^5$ is related to $T^5$ as in \snewt, by:
\eqn\snewtr{\eqalign{\widehat L_i&={l_p^3\over RL_i}\cr
M_s^2&={R^2L_1L_2L_3L_4L_5 \over l_p^9},}} where $R$ is the radius of
the longitudinal direction and $l_p$ is the eleven-dimensional
Planck-length.  Indeed, this gives the correct light-cone $X_G\times
T^5$ space-time from the moduli space of vacua (subject to the same
discussion about the situation at the quantum level as in 
\refs{\brs, \snewt}).  

In the limit of large $T^5$, this reduces to a slight variant of the
suggestion of \mrdgs\ outlined above.  The massless gauge group of the
$6d$ theory is given by \iibgg\ rather than $\prod _{\mu =0}^r
U(v_\mu)$; in addition, there are the $n_T=r$ tensor multiplets.  Upon
compactification, the tensor multiplets yield $U(1)^r$ gauge fields,
the same number which became massive because of the anomaly.  It is
thus tempting to conclude that, upon compactification, the tensor
multiplets simply give back the same $U(1)$ factors which became
massive in 6d because of the anomaly, giving back the original $\prod
_\mu U(Kn_\mu)$ theory in lower dimensions.  However, this does not
seem to be the case.  The difference is that the matter fields $\half
\oplus _{\mu \nu =0}^r a_{\mu \nu}(\fund _\mu , \overline{\fund}
_{\nu})$ were charged under the $U(1)^r$ which became massive because
of the 6d anomaly.  On the other hand, these matter fields are neutral
under the $U(1)^r$ which the tensor multiplets give back upon
compactification; the new $U(1)^r$ has no charged matter.  Taking the limit
of large $T^5$ in \snewtr\ thus yields a slight variant of the gauge theory
of \mrdgs.

Following \snewt, we similarly expect that the 6d string theory from
$SO(32)$ or $E_8\times E_8$ heterotic five-branes at a $X_G$
singularity, when compactified on $\widehat T^5$ (which depends on the
105 parameters in $SO(21,5,\IZ )\backslash SO(21,5)/(SO(21)\times
SO(5))$), gives a definition of $M$ theory on $X_G\times (T^5/\IZ
_2)\times \IR ^{1,1}$.

\bigskip
\centerline{{\bf Acknowledgments}}

I would like to thank J. Blum, D.R. Morrison, S. J. Rey, S. Sethi, 
E. Witten, and especially N. Seiberg for useful discussions.  This
work was supported by NSF PHY-9513835, the W.M. Keck Foundation, an
Alfred Sloan Foundation Fellowship, and the generosity of Martin and
Helen Chooljian.  The final stage of this work was also supported by
UCSD grant DOE-FG03-97ER40506.

\listrefs
\end